\documentclass[preprint,showpacs,preprintnumbers,amsmath,amssymb]{revtex4}

\usepackage{graphicx}
\usepackage{dcolumn}
\usepackage{bm}

\begin{document}
\allowdisplaybreaks{

\preprint{KEK-TH-1341}

\vspace*{25mm}

\title{Chiral Generations on Intersecting 5-branes\\
in Heterotic String Theory}

\author{Tetsuji Kimura}
\email{tetsuji@post.kek.jp}

\author{Shun'ya Mizoguchi}
\altaffiliation[Also at ]{Department of Particle and Nuclear Physics, The Graduate University for Advanced Studies.}
\email{mizoguch@post.kek.jp}
\affiliation{%
Theory Center, High Energy 
Accelerator Research Organization (KEK)\\
Tsukuba, Ibaraki 305-0801, Japan 
}
%

\date{Dec.8, 2009\\ \phantom{} }

\begin{abstract}
We show that there exist two {\bf 27} and one $\overline{\bf 27}$ of $E_6$, 
net one $D=4$, ${\cal N}=1$ chiral matter supermultiplet as zero modes 
localized on the intersection of two 5-branes in the 
$E_8 \times E_8$ heterotic 
string theory.  The smeared intersecting 5-brane solution is used 
via the standard embedding to construct a heterotic background, which provides, 
after a compactification of some of the transverse dimensions, a five-dimensional 
Randall-Sundrum II like
brane-world set-up in heterotic string theory. 
As a by-product, 
we present  a new proof of anomaly cancellation between those from the 
chiral matter and the anomaly inflow onto the brane
without small instanton. 
\end{abstract}

\pacs{11.25.Mj, 11.25.Wx, 11.30.Qc}
\maketitle

\section{\label{sec:Introduction}Introduction
}
How the Standard Model emerges in string theory is a long-standing question.
In early days of string theory, 
the heterotic string theory \cite{heterotic_string} 
was considered as a promising candidate for the fundamental theory which would 
provide a basis for model building. Its miraculous anomaly cancellation allows 
only two choices (that is, $E_8\times E_8$ and $SO(32)$) of a consistent gauge 
group, and in Calabi-Yau compactifications (including orbifold and other $1/4$ 
supersymmetric compactifications in a broad sense) there appear variety of 
four-dimensional supersymmetric standard-model-like theories with chiral generations. 
The problem is, however, that the number of such possible compactifications 
seems too large \cite{Susskind} to find natural necessity for our world to be 
as observed, despite the remarkable uniqueness of the original theory.

In the late last century, a conceptually different approach was proposed 
to realize a four-dimensional world by using D-branes 
in type II string theories. 
The key observation is that 
two intersecting D-branes can support chiral fermions at the intersection \cite{BDL}.
Since then many intersecting D-brane models have been built and discussed 
so far. We refer to the articles \cite{IntersectingDbraneModels} for a review of 
these developments. Also, inspired by the discovery of D-branes, 
brane-world models have also been extensively studied 
as a possible solution to the hierarchy problem and in terms of cosmological 
model building \cite{ADD, RS1, RS2}.

In this paper, we propose a new brane-world set-up for $E_6$ GUT 
model building by using intersecting 5-branes in {\em heterotic string theory}. 
The 5-branes in heterotic string theory are, of course, not D-branes. They are
NS5-branes \cite{SJR, Strominger}, and unlike D-branes, 
they are not described by open strings. 
What makes them hard to deal with is that, near the core of the solution, 
the geometry is not AdS but an infinite throat 
where the dilaton diverges linearly. Nevertheless, we can identify what low-energy 
excitations are on the brane by investigating zero modes of the
supergravity solution \cite{CHS}. 
It has been known for some time 
that on a symmetric 5-brane \cite{CHS} there are 30 $D=6$, ${\cal N}=1$ 
supermultiplet as zero modes in either of $E_8 \times E_8$ or $SO(32)$ 
heterotic string theory. In fact, as we explain in section \ref{sec:review_of_5brane}, 
they can be regarded as 
certain Nambu-Goldstone modes associated with various spontaneously broken 
symmetries of the theory. Therefore, we may expect that, as pions are effectively 
described 
by the chiral model without detailed knowledge of
QCD, the zero modes on the heterotic 5-branes may also provide enough
information for low-energy model building, even though their microscopic 
description (such as little string theory) is not fully understood. 
The existence of chiral zero modes is also consistent 
with the anomaly cancellation against an anomaly inflow from the bulk.

In order to examine the zero modes on the intersecting system, 
we first construct an intersecting 5-brane solution in the $E_8 \times E_8$ 
heterotic string theory 
by the so-called standard embedding in the known smeared intersecting 
NS5-brane solution of type II theories. We then study the zero modes of the 
relevant Dirac operator on this background. We show that there exist three 
localized chiral zero modes, two of which are in the {\bf 27} representation of $E_6$, 
and one in the $\overline{\bf 27}$ representation. They give rise to net {\em one} 
{\bf 27} of massless chiral fermions in the four-dimensional spacetime. 
Therefore, still being a toy model, 
this is the first example of a brane set-up in heterotic string theory  
that supports  
four-dimensional chiral matter fermions transforming as an $E_6$ gauge 
multiplet \footnote{This corrects the statement made in an earlier version of \cite{KM}, 
in which it was erroneously conjectured that the three supermultiplets 
would be of 
the same chirality.}.

There is a good reason why we should study NS5-branes 
as a set-up for particle-physics model building: They are T-dual to 
noncompact Calabi-Yau manifolds obtained by blowing up an 
isolated singularity \cite{T-dual}.
For instance, parallel NS5-branes are known to be T-dual to a multi-center
Taub-NUT, or an $A_n$ singularity which is obtained as a limit of a Taub-NUT. 
Similarly, a system of two intersecting 
5-branes is known to be T-dual to a deformed conifold 
\cite{conifold}.
Therefore, the intersecting 5-brane background in heterotic string theory 
may be regarded as a T-dual to a heterotic ``compactification'' on the 
deformed conifold.
While there are a variety of compact Calabi-Yau manifolds
with complicated structures, 
singularities may occur on a moduli space of {\em any} 
compact Calabi-Yau, and the local structure of a singularity is universal 
and can be simple, no matter what the rest of the manifold is.  
Therefore, the idea is that if a realistic GUT could be realized 
on such a singularity, it would mean that 
our universe is not just a coincidence, 
as {\em every} compact Calabi-Yau has a chance to realize the GUT 
on a part of it. 

%
%
%

This work is a first step toward a brane 
realization of a realistic $E_6$ GUT model in string theory.  
The remainder of this paper is organized as follows: 
In section II, we give a brief review of known 5-brane solutions in 
type II and heterotic supergravity theories. In section III, we present  
a new proof of anomaly cancellation between those from the chiral matter 
on the brane and the anomaly inflow into the brane in the $E_8\times E_8$ 
heterotic theory. In section IV, we construct an intersecting solution in the 
heterotic theory, and compute explicitly the zero modes
of the Dirac operator 
to find net $2-1=1$ set of chiral zero modes 
transforming as the {\bf 27} representation of $E_6$.
The last section is devoted to conclusions and discussion.

\section{\label{sec:review_of_5brane}Review of 5-brane Solutions 
in Heterotic String Theory
}
We will focus on the $E_8\times E_8$ heterotic string theory.
The string-frame bosonic supergravity Lagrangian is given,  to $O(\alpha')$
\cite{BdR,KY}, as
\begin{eqnarray}
{\cal L}&=&\frac1{2\kappa^2} \int d^{10}x  \sqrt{-g}e^{-2\phi}
\left\{
R(\omega)-\frac13 H_{MNP}H^{MNP}+4(\partial_M\phi)^2\right.
\nonumber\\
&& \left. \hspace{40mm}
-\alpha'\left(\frac{1}{30} \text{Tr} (F_{MN}F^{MN})
-R_{MNAB}(\omega_+) R^{MNBA}(\omega_+)\right)
\right\}.
\label{Lagrangian}
\end{eqnarray}
The convention we use in this paper is basically the one used in 
Callan-Harvey-Strominger's original paper  \cite{CHS}, and \cite{KY}, 
to which the reader is also referred for the comparison with 
other articles such as \cite{BdR}. 

As already seen in the above effective Lagrangian, particular combinations 
of the spin connection $\omega$ and the antisymmetric three-form $H$ play 
different roles in different places \cite{BdR,KY}. 
In (\ref{Lagrangian}), the $R^2$ term is 
the Riemann square made of the combination
\begin{eqnarray}
\omega_+&\equiv&\omega + H.
\end{eqnarray}
This combination also appears in the higher order terms in the effective 
action, and in the Bianchi identity for the $H$
field in the presence of flux:
\begin{eqnarray}
dH&=&\alpha'\left(
\text{tr} ( R(\omega_+)\wedge R(\omega_+)) -\frac1{30} \text{Tr}(F\wedge F)
\right). 
\label{Bianchi_H}
\end{eqnarray}
On the other hand, another combination 
\begin{eqnarray}
\omega_-&\equiv&\omega - H
\end{eqnarray}
is relevant for the lowest order SUSY variation of the gravitino:
\begin{eqnarray}
\delta \psi_M&=&\left(
\partial_M +\frac14 \omega_{-M}{}^{AB}\Gamma_{AB}
\right) \epsilon.
\end{eqnarray}
Finally, the Dirac operator of the gaugino equation of motion has a combination
$\omega-\frac13 H$ as will be seen in a moment.
The relations among the above three spin connections are discussed in 
\cite{TK0704}.

\subsection{The neutral solution}
In the absence of the nonabelian gauge field, the following configurations 
solve the leading order equations of motion:
\begin{eqnarray}
g_{ij}&=&\eta_{ij}~~~(i,j=0,1,\ldots,5),\nonumber\\
g_{\mu\nu}&=&e^{2\phi}\delta_{\mu\nu}~~~(\mu,\nu=6,\ldots,9),\nonumber\\
e^{2\phi}&=&e^{2\phi_0}+\frac {n\alpha'}{x^2},\nonumber\\
H_{\mu\nu\lambda}&=&-\epsilon_{\mu\nu\lambda}{}^{\rho}\partial_\rho \phi,
\label{neutral}
\end{eqnarray}
where 
\begin{equation}
x^2\equiv \sum_{\mu=6}^9(x^\mu)^2.
\end{equation}
$\epsilon^{\mu\nu\lambda\rho}$ 
is the (undensitized) completely antisymmetric tensor. 
All other components of $H$ vanish. 

The solution (\ref{neutral}) may be regarded as representing the NS5-branes 
stacked on top of each other
in both type IIA and type IIB theories. It has a nonzero axion charge
\begin{eqnarray} 
\frac1{2\pi^2}\int_{S^3} H &=& n\alpha'. 
\end{eqnarray} 
$n$ must be an integer. This is an everywhere smooth solution; $x=0$ is 
an apparent singularity as is verified by the coordinate transformation $t\equiv
\ln x^2$ \cite{CHS}. The scalar curvature and Riemann square (in the string frame) are 
both
everywhere finite:
\begin{eqnarray}
R
&=&\frac 6{n\alpha'}\cdot \frac{1}{(1+\frac{x^2}{\rho^2})^3},
\\
R_{ABCD}R^{ABCD}
&=&
\frac{12}{n^2{\alpha'}^2}\cdot
\frac{1+4\frac{ x^2}{\rho^2} + 8 (\frac{ x^2}{\rho^2})^2}{(1+\frac{ x^2}{\rho^2})^6},
\end{eqnarray}
where $\rho^2\equiv e^{-2\phi_0} n\alpha'$.
The supergravity analysis is trusted if the string coupling is small enough, and 
the metric varies slow enough:
\begin{eqnarray}
e^{2\phi} \ll 1, \hspace{5mm} R \ll {\alpha'}^{-1}.
\end{eqnarray}
They are satisfied if
\begin{eqnarray}
e^{2\phi_0} \ll 1, \hspace{5mm} \rho \ll |x|.
\end{eqnarray}

When considered in heterotic string theory later, the parameter $\rho$ 
corresponds to the size of the instanton. Therefore, a  small instanton 
means that the string coupling is everywhere large, and some nonperturbative 
phenomenon is known to occur \cite{small_instanton}.  
Even though $e^{-\phi_0}$ is large, the dilaton becomes large if one gets 
closer than the instanton size to the brane. However, 
a close relative of the symmetric 5-brane 
has been obtained \cite{GiddingsStrominger} as a certain double scaling limit of a non-extremal 
solution, and it is known to have, as a part of its near-horizon geometry, 
a two-dimensional black hole rather than a linear-dilaton throat geometry. 
CFT models inspired by this solution have been 
constructed \cite{noncompactGepner}. (The worldsheet approach for 
5-branes was originally mentioned in the second reference of \cite{SJR}.)

The zero modes on this solution are a six-dimensional 
{\em chiral} $(2,0)$ matter supermultiplet in the IIA case, and {\em nonchiral} $(1,1)$ 
supermultiplet in the IIB case \cite{CHS}. This flip of the chirality may be 
understood as a consequence of the T-duality to the ADE singularities.

\subsection{The symmetric solution}
Next we include a nonabelian gauge field in heterotic string theory.
It is well known that in order for the anomaly cancellation mechanism 
to work, the Bianchi identity must be modified as we saw in (\ref{Bianchi_H}) as to
\begin{eqnarray}
dH&=&\alpha'\left(\text{tr}( R(\omega_+)\wedge R(\omega_+)) 
-\frac1{30} \text{Tr}(F\wedge F)
\right),
\end{eqnarray}
where Tr is the trace in the adjoint representation of $E_8 \times E_8$ or $SO(32)$.

Since the neutral solution satisfies $dH=0$ except at the brane position $x^\mu=0$
where the magnetic 5-brane charge is located, it remains a solution in heterotic 
theory only if the right hand side vanishes. The most common way to achieve this 
is to set $\omega_+$ equal to the gauge connection $A$. This may be called the 
``standard'' embedding, but the point is that, in the presence of nonzero $H$ flux, 
what is embedded in the gauge connection is not simply the spin connection 
$\omega$, but the particular combination $\omega+H$.
What is nice about this embedding is that some corrections of the supersymmetry 
variations to higher orders in $\alpha'$ vanish \cite{BdR}.   

The spin connections computed in the neutral background (\ref{neutral}) are 
found to be
\begin{eqnarray}
\omega_{\mu}{}^{\alpha}{}_{\beta}&=&
(\delta_\mu^\alpha \delta_\beta^\nu-\delta_{\mu\beta}\delta^{\alpha\nu})
\partial_\nu\phi.
\end{eqnarray}
All other components vanish.
The $H$ field is written as 
\footnote{In these expressions no vielbein appears because the metric is 
diagonal and the scale factors cancel in the present case. }
\begin{eqnarray}
H_{\mu}{}^{\alpha}{}_{\beta}&=&
-\epsilon^{\alpha}{}_{\beta}{}^{\gamma\delta} \, \delta_{\mu\gamma} \delta_\delta^{~\nu}
\partial_\nu\phi.
\end{eqnarray}
Therefore 
$\omega_+$ is given by
\begin{eqnarray}
\omega_{+\mu}{}^{\alpha\beta}&\equiv&(\omega+H)_{\mu}{}^{\alpha\beta}
\nonumber\\
&=&2\rho^2\sigma^{\alpha\beta}{}_{\mu\lambda} \cdot
\frac{x^\lambda}{x^2(x^2+\rho^2)},
\end{eqnarray}
where 
\begin{eqnarray}
\sigma^{\alpha\beta}{}_{\mu\lambda}
&\equiv&\delta_\mu^\alpha \delta_\lambda^\beta
-\frac12\epsilon^{\alpha\beta}{}_{\gamma\delta} \, \delta_\mu^\gamma \delta_\lambda^\delta,
\\
\rho^2&\equiv&e^{-2\phi_0}n\alpha'.
\end{eqnarray}
The tensor $\sigma^{\alpha\beta}{}_{\mu\lambda}$ is anti-self-dual:
\begin{eqnarray}
\frac12 \epsilon^{\alpha\beta}{}_{\gamma\delta} \, \sigma^{\gamma\delta}{}_{\mu\nu}&=& 
-\sigma^{\alpha\beta}{}_{\mu\nu}.
\end{eqnarray}
Thus the $SO(4)$ connection $\omega_+$ is anti-self-dual, which means that 
the structure group of the bundle is reduced to $SU(2)$. 
We then identify
\begin{eqnarray}
A_\mu^{\alpha\beta}&=&\omega_{+\mu}{}^{\alpha\beta},
\label{embeddedSU(2)A}
\end{eqnarray}
and assume all other components to be zero.
This is the symmetric solution \cite{CHS}. 
In this way, a part of the gauge connection  
acquires a nonzero background in an $SU(2)$ subalgebra of $E_8 \times E_8$.

This is a supersymmetric configuration; 
a different combination 
\begin{eqnarray}
\omega_{-\mu}{}^{\alpha\beta}&\equiv&(\omega-H)_{\mu}{}^{\alpha\beta}
\nonumber\\
&=&2\rho^2\hat\sigma^{\alpha\beta}{}_{\mu\lambda} \cdot
\frac{x^\lambda}{x^2(x^2+\rho^2)},
\\
\hat\sigma^{\alpha\beta}{}_{\mu\lambda}
&\equiv&\delta_\mu^\alpha \delta_\lambda^\beta
+\frac12\epsilon^{\alpha\beta}{}_{\gamma\delta} \, \delta_\mu^\gamma \delta_\lambda^\delta
\nonumber
\end{eqnarray}
is a {\em self-dual} connection, and hence belongs also to a (different) $SU(2)$ 
subalgebra of $SO(4)$. 
This ensures that there is a Killing spinor 
for the gravitino SUSY variation 
\begin{eqnarray}
\delta \psi_M&=&\left(
\partial_M +\frac14 \omega_{-M}{}^{AB}\Gamma_{AB}
\right) \epsilon.
\label{SUSYgravitino}
\end{eqnarray}
On the other hand, the gaugino SUSY variation reads
\begin{eqnarray}
\delta \chi^{\alpha\beta}&=&-\frac14 \Gamma^{MN}F_{MN}^{\alpha\beta}\epsilon,
\label{SUSYgaugino}
\end{eqnarray}
where the $SO(4)$ matrix indices $\alpha,\beta$ are now understood as the $SU(2)$ 
gauge indices. The field strength $F_{MN}^{\alpha\beta}$ involves
the connection $\omega_+$ due to the embedding, and not $\omega_-$. 
However, there is a following identity between the Riemann tensor made of the connection
$\omega_+$ and that made of $\omega_-$:
\begin{eqnarray}
R(\omega_+)_{MNPQ}&=&
R(\omega_-)_{PQMN}+ (dH)_{MNPQ}.
\end{eqnarray}
Therefore, in the background where $dH$ vanishes, the gaugino variation 
(\ref{SUSYgaugino}) amounts to
\begin{eqnarray}
\delta \chi^{\alpha\beta}&=&-\frac14 \Gamma^{\mu\nu}F_{\mu\nu}^{\alpha\beta}\epsilon
\nonumber\\
&=&-\frac14 \Gamma^{\mu\nu}R(\omega_+)_{\mu\nu}{}^{\alpha\beta}\epsilon
\nonumber\\
&=&-\frac14 \Gamma^{\gamma\delta}R(\omega_-)^{\alpha\beta}{}_{\gamma\delta}\epsilon.
\end{eqnarray}
Thus the Killing spinor $\epsilon$ for the gravitino variation (\ref{SUSYgravitino}) is 
automatically the Killing spinor for the gaugino variation (\ref{SUSYgaugino}). 
(The dilatino variation equation must be checked separately.)
The $SU(2)$ gauge connection $A_\mu^{\alpha\beta}$ (\ref{embeddedSU(2)A}) 
satisfies the lowest-order equation of motion
\begin{eqnarray}
\partial_\nu\left(\sqrt{-g} e^{-2\phi} F^{\mu\nu\alpha\beta}\right)
+\sqrt{-g} e^{-2\phi}
\left({[}A_\nu,~F^{\mu\nu} {]}^{\alpha\beta}
-H^{\mu\nu\rho} F_{\nu\rho}^{\alpha\beta}
\right)&=&0 \label{F_eom}
\end{eqnarray}
as expected.

\subsection{Zero modes on the symmetric 5-brane}
Let us consider zero modes existing on the symmetric 5-brane solution (\ref{neutral}) 
with (\ref{embeddedSU(2)A}) \cite{CHS}. The obvious bosonic zero modes are the four translation 
moduli, and the instanton size $\rho$ modulus. Besides, there are other zero modes 
coming from infinitesimal global gauge rotations of the instanton:
By construction, the gauge fields have nonzero vacuum expectation values in 
the four-dimensional space transverse to the 5-brane. They belong to an $SU(2)$ 
subalgebra of one of $E_8$. The centralizer of $SU(2)$ in $E_8$ is $E_7$,
and the adjoint {\bf 248} is decomposed into a sum of representations of $E_7\times SU(2)$
as
\begin{eqnarray}
{\bf 248}&=&({\bf 133},{\bf 1}) \oplus ({\bf 56},{\bf 2}) \oplus ({\bf 1},{\bf 3}).
\end{eqnarray}    
{\bf 133}, the adjoint of $E_7$, does nothing on the $SU(2)$ background, while 
the other $56\times 2 + 1\times 3 = 115$ generators rotate the background, 
and hence give rise to zero modes. Thus, in all, there are $4+1+115=120$ bosonic 
zero modes on this background. Since the symmetric solution is half BPS, 
they together with their superpartners constitute  30 $D=6$, ${\cal N}=1$ hypermultiplets. 
The existence of 
the fermionic zero modes have also been confirmed by the index theorem 
\cite{Bellisai}.

These zero modes can be regarded as Nambu-Goldstone modes associated with 
various spontaneously broken symmetries of the theory \cite{HughesPolchinski}.
Indeed, the four position moduli above are the Nambu-Goldstone modes 
coming from the spontaneous broken translational invariance due to 
the presence of the 5-brane. The size modulus corresponds to the broken 
scale invariance. The remaining $115$ moduli are also 
thought of as coming from how the $SU(2)$ subalgebra is embedded in 
the whole $E_8$ Lie algebra; by ``standard embedding'' we mean we choose 
{\em some} $SU(2)$ in $E_8$ and set the gauge connection for {\em this}
$SU(2)$ to be equal to the (generalized) spin connection. But the choice of such 
an $SU(2)$ is arbitrary, and the original $E_8$ symmetry is spontaneously 
broken.
Incidentally, this way of counting reproduces the correct instanton-number
dependence of the dimensions of instanton moduli in flat space, for all gauge 
groups, obtained by the index theorem \cite{BCGW}.   

But there is a puzzle here: Why aren't they absorbed into the gauge bosons 
by the Higgs mechanism? The gauge bosons in the transverse dimensions 
can be viewed  as adjoint Higgs fields from the brane, and the standard embedding 
amounts to giving vev's to these Higgs fields. Then small fluctuations around 
the vev's are Nambu-Goldstone modes, which are completely 
gauged away to leave, in ordinary gauge theories, a Proca Lagrangian for 
massive vector fields. This is the standard Higgs mechanism in the textbook, and 
it is interpreted to mean that the Nambu-Goldstone modes are ``eaten'' by
the gauge bosons to be their longitudinal degrees of freedom.  
So why are there such extra zero-mode degrees of freedom left on the brane, other than those 
used as a part of massive vector bosons in the bulk?   

The resolution to this problem lies 
\footnote{We are grateful to H.~Kawai, H.~Kunitomo and N.~Ohta for 
discussions on this issue.} in the apparent breakdown of the gauge 
invariance due to the Green-Schwarz counterterm $BX_8 \sim -dB X_7$. 
In eliminating the small fluctuations around the vev, 
both $B$ and $X_7$ also get transformed by the gauge transformation.
The contribution from the variation of $B$ is compensated by 
the one-loop anomaly in the bulk \cite{GSmechanism},    
while that from $X_7$ vanishes if there are no magnetic source of the 
$B$ field $d^2B=0$. In the present case, however, there {\em is} 
such a source $d^2B\propto \delta^4(x^\mu)$, 
and therefore the gauge variation of $X_7$ gives rise to a change of 
the field configurations on the brane. Thus gauge transformations can {\em not 
completely} eliminate the fluctuations of the ``Higgs'', but local fluctuations  
are left on the brane
\footnote{In contrast, 
zero modes coming from an abelian gauge field in other theories 
(such as type IIA theory
\cite{CHS} and $D=5$ supergravity \cite{MizoguchiOhta})
are not pure gauge rotations.}.

This phenomenon is known as anomaly inflow \cite{anomaly_inflow}, 
and the change of the brane action is cancelled by, again, the one-loop 
effect of chiral fermions on the brane, which are the superpartners of the 
bosonic zero modes. The gauge invariance of the total quantum action 
is thus restored. In the next section, we will show the precise arithmetic of the 
cancellation.

\section{Anomaly Inflow and Cancellation}
We will show that the 30 hypermultiplets, 28  (=56 half-hypermultiplets) 
in the {\bf 56} representation of $E_7$ and two singlets (=4 half-hypermultiplets), 
precisely cancel 
the inflows of the tangent bundle, $E_7$ gauge and mixed anomalies via 
the Green-Schwarz mechanism.  
The cancellation of anomalies on the gauge 5-brane \cite{CHS} in heterotic 
string theory was already discussed in \cite{BlumHarvey}. Here we give a somewhat 
different proof of cancellation than theirs in the case of the $E_8\times E_8$
symmetric 5-brane. Although they should be basically the same, ours is closely 
parallel to Mourad \cite{Mourad} and appears to be simpler. 
In particular, we do not need to 
consider any current at infinity. 
We ignore the normal bundle connection 
and write out only terms consisting of the tangent bundle and gauge 
connections.

The relevant anomaly polynomials are    
\begin{eqnarray}
I_8^{singlet}&=&\left.\frac12 \hat{A}(T\Sigma)\right|_8 \times 4\nonumber\\
&=&\frac 2{5760}(-4 p_2 + 7 p_1^2),\\
I_8^{\bf 56}&=&\left.\frac12 \hat{A}(T\Sigma)\mbox{tr}_{\bf 56}e^{iF} 
\right|_8 \nonumber\\
&=&\frac {28}{5760}(-4 p_2 + 7 p_1^2)
+\frac1{96}p_1^2\mbox{tr}_{\bf 56}F^2 
+\frac1{48}\mbox{tr}_{\bf 56}F^4, 
\end{eqnarray}
%
%
and
\begin{eqnarray} 
X_8&=&\frac1{24}
\left(
\frac18\mbox{tr}R^4
+\frac1{32}(\mbox{tr}R^2)^2
-\frac1{240}\mbox{tr}R^2 \mbox{Tr}_{\bf 248}F^2
+\frac1{24}\mbox{Tr}_{\bf 248}F^4
-\frac1{7200}(\mbox{Tr}_{\bf 248}F^2)^2
\right)\nonumber\\
&=&
\frac {1}{192}(-4 p_2 + 7 p_1^2)
+\frac1{2880}p_1^2\mbox{Tr}_{\bf 248}F^2 
+\frac1{576}\mbox{Tr}_{\bf 248}F^4
-\frac1{24\cdot 7200}(\mbox{Tr}_{\bf 248}F^2)^2.
\end{eqnarray}
Since the gauge symmetry is broken from $E_8$ to $E_7$, 
we rewrite the traces in the representations of $E_8$ to those 
of $E_7$. The following formulas are useful \cite{Erler}: 
\begin{eqnarray}
\mbox{Tr}_{\bf 248}F^2&=&(\mbox{Tr}_{\bf 133}+2\mbox{tr}_{\bf 56})F^2\nonumber\\
&=&5\mbox{tr}_{\bf 56}F^2,\\
\mbox{Tr}_{\bf 248}F^4&=&(\mbox{Tr}_{\bf 133}+2\mbox{tr}_{\bf 56})F^4\nonumber\\
&=&\frac14(\mbox{tr}_{\bf 56}F^2)^2.
\end{eqnarray}
\begin{eqnarray}
\mbox{Tr}_{\bf 133}F^2&=&3\mbox{tr}_{\bf 56}F^2,\nonumber\\
\mbox{Tr}_{\bf 133}F^4&=&\frac16(\mbox{tr}_{\bf 56}F^2)^2,\nonumber\\
\mbox{tr}_{\bf 56}F^4&=&\frac1{24}(\mbox{tr}_{\bf 56}F^2)^2.
\end{eqnarray}
Therefore
\begin{eqnarray}
X_8
&=&
\frac {1}{192}(-4 p_2 + 7 p_1^2)
+\frac1{24^2}p_1^2\mbox{tr}_{\bf 56}F^2 
+\frac1{4\cdot24^2}(\mbox{tr}_{\bf 56}F^2)^2
-\frac1{12\cdot 24^2}(\mbox{tr}_{\bf 56}F^2)^2.
\end{eqnarray}
They add up to
\begin{eqnarray}
I_8^{singlet}+I_8^{\bf 56}-X_8
&=&\frac1{3\cdot 24^2}
(3p_1 +\mbox{tr}_{\bf 56}F^2)
(12p_1 +\mbox{tr}_{\bf 56}F^2).
\label{I+I+X}
\end{eqnarray}

Note that the number (thirty) of hypermultiplets is precisely the one 
which can cancel out the $p_2$ term, otherwise the sum of anomalies 
does not factorize and the Green-Schwarz mechanism does not 
apply. 
Since
\begin{eqnarray}
12p_1 +\mbox{tr}_{\bf 56}F^2
&=&
6\left(
-\mbox{tr}R^2 +\frac1{30}\mbox{Tr}_{\bf 248}F^2
\right),
\end{eqnarray}
which is proportional to the anomalous part of the heterotic Bianchi 
identity, the sum (\ref{I+I+X}) is cancelled by introducing a Green-Schwarz 
counterterm on the brane as in \cite{Mourad}.

\section{\label{sec:intersecting_5branes}Intersecting 5-branes  
in Heterotic String Theory
}
\subsection{Intersecting neutral 5-branes}
We will now consider a system of two intersecting NS5-branes. 
We start with the neutral smeared solution \cite{intersecting_solutions}:
\begin{eqnarray}
ds^2&=&\sum_{i,j=0,7,8,9}\eta_{ij}dx^i dx^j
+h(x^1)^2\sum_{\mu,\nu=1,2}\delta_{\mu\nu}dx^\mu dx^\nu
+h(x^1)\sum_{\mu,\nu=3,4,5,6}\delta_{\mu\nu}dx^\mu dx^\nu,
\nonumber\\
e^{2\phi}&=&h(x^1)^2,\nonumber\\
H_{\mu\nu\lambda}&=&\left\{
\begin{array}{cl}
\frac{h'}2 ~(=\frac{\xi|x^1|'}2)&\mbox{if $(\mu,\nu,\lambda)=(2,3,4)$,$(2,5,6)$ or their even permutation},\\
-\frac{h'}2~(=-\frac{\xi|x^1|'}2)&\mbox{if $(\mu,\nu,\lambda)=(2,4,3)$,$(2,6,5)$ or their even permutation},\\
0&\mbox{otherwise,}
\end{array}
\right.
\label{intersecting_neutral}
\end{eqnarray}
where
\begin{equation}
h(x^1)=h_0+\xi |x^1|.
\end{equation}
All other components of $H_{MNL}$ vanish. $h_0$ and $\xi$ are real constants. 
The prime ${}'$ denotes 
the differentiation with respect to $x^1$, and $|x^1|'$ is therefore a step function.
This is a solution to equations of motion of the leading-order NSNS-sector 
Lagrangian in type II theories:
\begin{eqnarray}
{\cal L}_{NS}&=&
\frac1{2\kappa^2} \int d^{10}x  \sqrt{-g}e^{-2\phi}
\left(
R(\omega)-\frac13 H_{MNP}H^{MNP}+4(\partial_M\phi)^2
\right).
\label{LagNSNS}
\end{eqnarray}
The solution describes a pair of intersecting NS5-branes 
stretching in dimensions as shown in TABLE \ref{tab:table1}. 
These branes are delocalized in the $x^2,x^3,x^4,x^5$ and $x^6$ directions.
Consequently, the solution depends only on $x^1$, and
hence the name ``smeared solution''.

\begin{table}
\caption{\label{tab:table1}\sl
Dimensions in which the 5-branes stretch. 
}
\begin{ruledtabular}
\begin{tabular}{lcccccccccc}
&0
&1
&2&3&4&5&6&7&8&9\\
\hline
5-brane1 & $\bigcirc$ & &&
&&$\bigcirc$&$\bigcirc$&$\bigcirc$&$\bigcirc$&$\bigcirc$\\
5-brane2 & $\bigcirc$ &&
&$\bigcirc$&$\bigcirc$&
&&$\bigcirc$&$\bigcirc$&$\bigcirc$\\
\end{tabular}
\end{ruledtabular}
\end{table}

\subsection{Brane tension and  the harmonic function}
The coefficient $\xi$ in the definition of the harmonic function $h(x^1)$ 
is related to the tension of the brane. To see this, let us consider 
Einstein's equation in the Einstein frame:
\begin{eqnarray}
&&({R_E})_A{}^{B}-\frac12 \delta_A{}^{B} {R_E} - T_A{}^{B}=\left\{
\begin{array}{ll}
0&(A,B=1,2),\\
+\frac{h''}{2h^{\frac{5}{2}}}&(A,B=3,4,5,6),\\
+\frac{h''}{h^{\frac{5}{2}}}&(A,B=0,7,8,9),
\end{array}
\right.
\label{Einstein's_eq}
\\
&&-T_A{}^{B}=-\frac12 \partial_A\phi \partial^B\phi
-e^{-\phi}H_{ACD}H^{BCD}
+\frac12\delta_A^B\left(
\frac12(\partial\phi)^2 +\frac13 e^{-\phi}H^2
\right).
\end{eqnarray}
The fact that the right hand side of (\ref{Einstein's_eq}) does not vanish 
implies that the action must include $\delta$-function like brane-energy terms:
\begin{eqnarray}
{\cal L}_E&=&\frac1{2\kappa^2}\sqrt{-G}R_E+{\cal L}_E(\phi,H) 
-V\sqrt{-G_{\text{\scriptsize 5-brane1}}}\prod_{\mu'=1,2,3,4}\delta(x^{\mu'})
\nonumber\\
&&~~~~~~~~~~~~~~~~~~~~~~~~~~~~~~~
-V\sqrt{-G_{\text{\scriptsize 5-brane2}}}\prod_{\mu''=1,2,5,6}\delta(x^{\mu''}),
\end{eqnarray}
where
$V$
is the brane tension. ${\cal L}_E(\phi,H)$ is the Lagrangian for the $\phi$ and
$H$ fields in the Einstein frame, which contributes to the energy-momentum tensor 
$T_A{}^{B}$ in (\ref{Einstein's_eq}).
The brane metrics are defined as
\begin{eqnarray}
{(G_{\text{\scriptsize 5-brane1}})}_{i'j'}
&=&G_{i'j'}(x^{\mu'}=0)~~~(i',j'=0,5,6,7,8,9;~\mu'=1,2,3,4),\nonumber\\
{(G_{\text{\scriptsize 5-brane2}})}_{i''j''}
&=&G_{i''j''}(x^{\mu''}=0)~~~(i'',j''=0,3,4,7,8,9;~\mu''=1,2,5,6).
\end{eqnarray}
What we see here is a no-cosmological-constant analogue of the
Randall-Sundrum (RS) models \cite{RS1, RS2}, 
and the intersecting nature of the solution is reflected in the two different 
brane-energy terms. After delocalizations:
\begin{eqnarray}
\prod_{\mu'=2,3,4}\delta(x^{\mu'})&\rightarrow&1,\nonumber\\
\prod_{\mu''=2,5,6}\delta(x^{\mu''})&\rightarrow&1,
\end{eqnarray}
the inclusion of these terms matches (\ref{Einstein's_eq}) if
\begin{eqnarray}
\xi&=&-\kappa^2 V h_0^{\frac52}.
\label{xi}
\end{eqnarray}

Since $e^\phi=h(x^1)$, the sign of $\xi$ strongly affects the dilaton 
profile. (If $\xi=0$, the solution is reduced to a flat Minkowski space.) 
We consider the following two cases separately:

\begin{figure}[b]
\includegraphics[height=0.2\textheight]{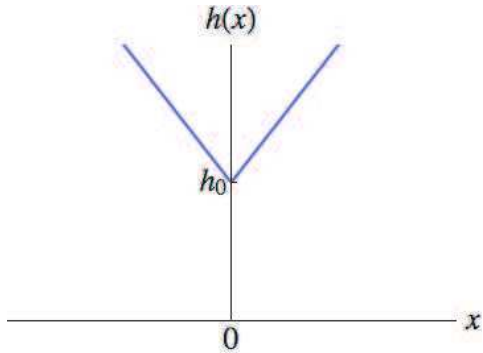}
\caption{\label{positive_xi_a}\sl 
$h(x)$ with $\xi>0$. 
The brane tension is negative.  Also, 
the string coupling becomes stronger as one goes away 
from the branes.}
\end{figure}
\begin{figure}[h]
{\renewcommand{\arraystretch}{.5}
\begin{tabular}{c@{\hspace{10mm}}c}
\includegraphics[height=0.2\textheight]{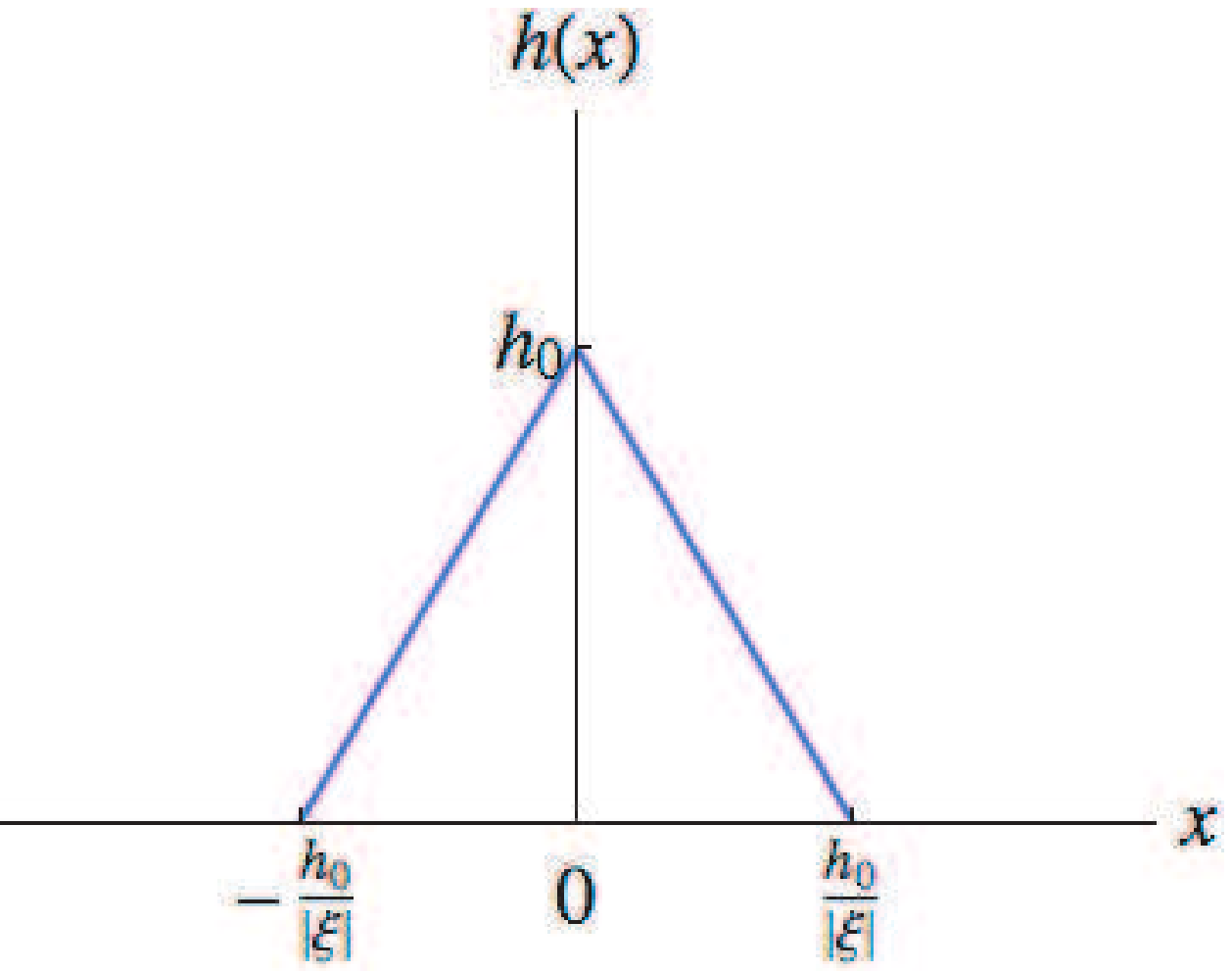}
&
\raisebox{1.5ex}{
\includegraphics[height=0.19\textheight]{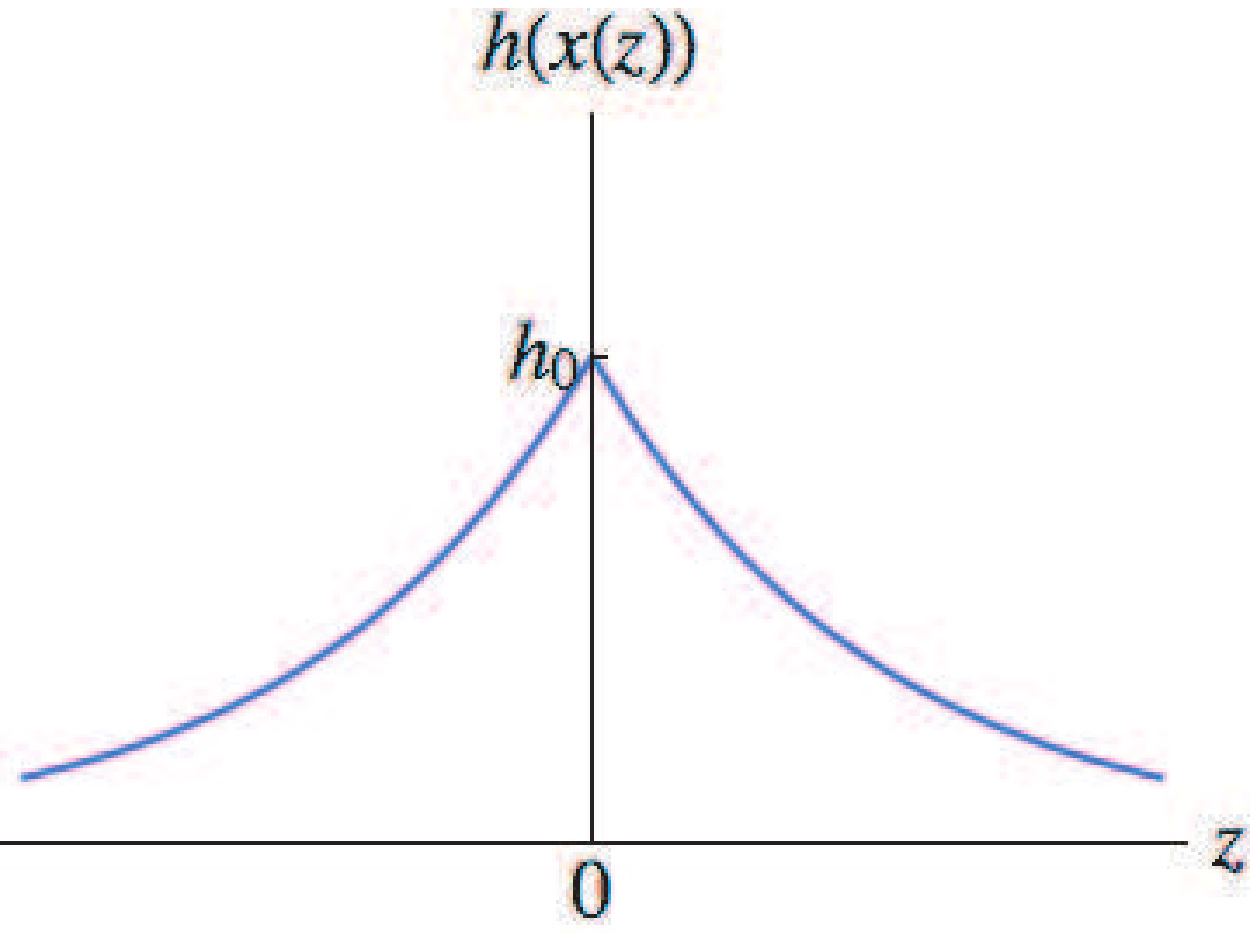}}
\\
(a) & (b)
\end{tabular}
}
\caption{
\label{negative_xi}\sl
$h(x)$ with $\xi<0$.
(a)
The brane has a positive tension. The string coupling  
decreases linearly from a positive value $h_0$, to necessarily cross 
the $x^1$ axis, where the string coupling becomes zero.  We 
identify this point as the ``end of the world''.
(b) 
By a change of the coordinate the points 
$x=\pm\frac{h_0}{|\xi|}$ are mapped to $z=\pm\infty$. The profile of 
$h(x(z))$ becomes similar to the warp factor of the RS II model.}
\end{figure}

If $\xi >0$ as in FIG. \ref{positive_xi_a},  the brane tension is negative. It is doubtful whether such 
an object may consistently exist in heterotic string theory. Also, 
if $\xi >0$, the string coupling becomes stronger as one goes away 
from the branes, which is puzzling. Thus we consider another option.

If $\xi<0$ as in FIG. \ref{negative_xi}, the brane has a positive tension. The function $h(x^1)$, 
and hence the string coupling, is now convex upwards in $x^1$. It 
decreases linearly from a positive value $h_0$, to necessarily cross 
the $x^1$ axis, where the string coupling becomes zero. Beyond that 
point, $h(x^1)$ becomes negative, which is inconsistent. Thus we 
identify this point as the ``end of the world''; one can 
send this point infinitely far away \footnote{Of course, this is just a change of a 
coordinate, and hence does not change the geodesic distance. Also it is 
not smooth at $x^1=0$ ($z=0$), and gives rise to an extra 
delta function in the second derivative.} by the coordinate transformation
\begin{eqnarray}
z&=&-\mbox{sign($x^1$)}\log \frac{h(x^1)}{h_0},
\label{z}
\end{eqnarray}  
where $z$ is the new coordinate. Then the function $h(x^1)$, which is 
the string coupling and a typical warp factor for the relatively transverse 
dimensions, is expressed simply as
\begin{eqnarray}
h&=&h_0~e^{-|z|}.
\end{eqnarray}

Apparently, this looks similar to the RS II model \cite{RS2}, but 
there are the following differences:
The first is that we have no bulk cosmological constant. Instead, 
we have the dilaton and axion fields (and also the nonabelian gauge fields 
after the standard embedding) which balance gravity. 
Secondly, as we see in a moment, there exist chiral zero modes on the branes,
which are in the ${\bf 27}$ representation of $E_6$. This is not an assumption 
but a logical consequence of string theory. 
The final difference is in the warp factor. Unlike the RS models, 
our four-dimensional metric is not warped at all in the 
string frame \footnote{More curiously, 
although the branes have a positive tension 
as we have derived (\ref{xi}), 
the 4D metric is inversely warped 
(like near the negative tension brane in the RS I model \cite{RS1})
in the Einstein frame.}.
It would be interesting to examine whether gravity or gauge field is localized, 
but in this paper we will focus only on the localization of chiral fermions.

\subsection{Intersecting 5-branes in heterotic string theory}
We now construct an intersecting solution in the $E_8 \times E_8$ 
heterotic string theory by the standard embedding, similarly to 
the previous parallel brane case.

The (generalized) spin connections 
of the neutral intersecting background are computed as
\begin{eqnarray}
(\omega \pm H)_{\mu=1}{}^{\alpha}{}_{\beta}
&=&0,
\nonumber\\
(\omega \pm H)_{\mu=2}{}^{\alpha}{}_{\beta}
&=&\frac{h'}h
\left(
\renewcommand{\arraystretch}{.7}
\begin{array}{cccccc}
 & -1  &&&&   \\
1 &   &&&&   \\
 &&&\pm \frac12   &&\\
 &&  \mp \frac12 &&&\\
&&&&&\pm \frac12\\
 &&&&  \mp \frac12 &
\end{array}
\right),
\nonumber\\
(\omega \pm H)_{\mu=3}{}^{\alpha}{}_{\beta}
&=&\frac{h'}{2h^{\frac32}}
\left(
\renewcommand{\arraystretch}{.7}
\begin{array}{cccccc}
 &   &-1&&&   \\
 &   &  &\mp 1&&   \\
~ 1&&& &&\\
 & \pm 1&&&&\\
&&&&~~~&\\
 &&&&&~~~
\end{array}
\right),
\nonumber\\
(\omega \pm H)_{\mu=4}{}^{\alpha}{}_{\beta}
&=&\frac{h'}{2h^{\frac32}}
\left(
\renewcommand{\arraystretch}{.7}
\begin{array}{cccccc}
 &   &&-1&&   \\
 &   &\pm1  &&&   \\
&\mp 1&& &&\\
~ 1 &&&&&\\
&&&&~~~&\\
 &&&&&~~~
\end{array}
\right),
\nonumber\\
(\omega \pm H)_{\mu=5}{}^{\alpha}{}_{\beta}
&=&\frac{h'}{2h^{\frac32}}
\left(
\renewcommand{\arraystretch}{.7}
\begin{array}{cccccc}
 &   &&&-1&   \\
 &   &  &&&\mp 1   \\
&&~~~&&& \\
 &&& ~~~&&\\
~ 1&&&&&\\
 &\pm 1&&&&
\end{array}
\right),
\nonumber\\
(\omega \pm H)_{\mu=6}{}^{\alpha}{}_{\beta}
&=&\frac{h'}{2h^{\frac32}}
\left(
\renewcommand{\arraystretch}{.7}
\begin{array}{cccccc}
 &   &&&&-1  \\
 &   &&&\pm1  &   \\
&&&~~~&& \\
&&~~~&&&\\
&\mp 1&&&&\\
 ~ 1&&&&&
\end{array}
\right).
\label{omega+-H}
\end{eqnarray}
The gauge connections are obtained by identifying
\begin{eqnarray}
A_\mu^{\alpha\beta}&=&(\omega+ H)_{\mu}^{~~\alpha\beta}.
\end{eqnarray}
The result is
\begin{eqnarray}
A_{\mu=1}^{~~\alpha\beta}&=&0,\nonumber\\
A_{\mu=2}^{~~\alpha\beta}&=&
\frac{h'}{h}
\left(
\renewcommand{\arraystretch}{.7}
\begin{array}{ccc}
-s &&\\
&\frac12 s &\\
&& \frac12 s
\end{array}
\right)
=
\frac{h'}{h}
\left(
- \frac{3\lambda_3 + \sqrt{3} \lambda_8}4
\right)
\otimes s ,\nonumber\\
A_{\mu=3}^{~~\alpha\beta}&=&
\frac{h'}{2h^{\frac 32}}
\left(
\renewcommand{\arraystretch}{.7}
\begin{array}{ccc}
&-{\bf 1}&\\
{\bf 1}& &\\
&&~~~
\end{array}
\right)
=
\frac{h'}{2h^{\frac 32}}
\left(
-i \lambda_2\right)
\otimes {\bf 1},\nonumber\\
A_{\mu=4}^{~~\alpha\beta}&=&
\frac{h'}{2h^{\frac 32}}
\left(
\renewcommand{\arraystretch}{.7}
\begin{array}{ccc}
&-s&\\
-s& &\\
&&~~~
\end{array}
\right)
=
\frac{h'}{2h^{\frac 32}}
\left(
- \lambda_1\right)
\otimes s,\nonumber\\
A_{\mu=5}^{~~\alpha\beta}&=&
\frac{h'}{2h^{\frac 32}}
\left(
\renewcommand{\arraystretch}{.7}
\begin{array}{ccc}
&&-{\bf 1}\\
&~~~&\\
~{\bf 1}&&~~~
\end{array}
\right)
=
\frac{h'}{2h^{\frac 32}}
\left(
-i \lambda_5\right)
\otimes {\bf 1},\nonumber\\
A_{\mu=6}^{~~\alpha\beta}&=&
\frac{h'}{2h^{\frac 32}}
\left(
\renewcommand{\arraystretch}{.7}
\begin{array}{ccc}
&&-s\\
&~~&\\
-s&&
\end{array}
\right)
=
\frac{h'}{2h^{\frac 32}}
\left(
- \lambda_4\right)
\otimes s,
\label{A_SU(3)}
\end{eqnarray}
where
$\lambda_i$'s ($i=1,\ldots,8$) are the Gell-Mann matrices and   
${\bf 1} \equiv\left(
\renewcommand{\arraystretch}{.7}
\begin{array}{cc}
1&\\~~&~~1
\end{array}
\right)$, 
$s \equiv i\sigma_2=\left(
\renewcommand{\arraystretch}{.7}
\begin{array}{cc}
&~1\\-1&
\end{array}
\right)$.

The explicit expressions (\ref{omega+-H}) show that both of $\omega_\pm$
are $SU(3)$ connections. As we did in section II for the symmetric 5-brane, we
have embedded $\omega_+$ into the gauge connection $A$. Then the Bianchi identity
is reduced to $dH=0$, and the solution (\ref{intersecting_neutral}) 
is consistent with it. This time a certain $SU(3)$ piece of the $E_8(\times E_8)$ 
gauge connection is given a nonzero expectation value. On the other hand, the fact that
$\omega_-\in SU(3)$ implies that the Killing spinor equations for the gravitino variation
(\ref{SUSYgravitino}) as well as, as explained before, the gaugino variation 
(\ref{SUSYgravitino}) have a common single Killing spinor. 
It can be checked that this also satisfies the equation for the dilatino SUSY variation 
to lowest order:
\begin{eqnarray}
\delta\lambda&=&\left(
-\frac14 \Gamma^M \partial_M \phi +\frac1{24}\Gamma^{MNP}H_{MNP}
\right)\epsilon~=0.
\end{eqnarray}
Thus the background (\ref{intersecting_neutral}) together with (\ref{A_SU(3)}) preserve
1/4 of supersymmetries. It also satisfies the equations of motion (\ref{F_eom})
as it should.  

\subsection{Zero modes as Nambu-Goldstone modes on the intersecting 5-branes}
In the previous subsection we have constructed a smeared solution which 
describes intersecting 5-branes in the $E_8\times E_8$ heterotic string theory 
to leading order in $\alpha'$, via the standard embedding, similarly to 
the way we obtain the symmetric 5-brane. In that case, the connection $\omega_+$ 
embedded was in $SU(2)$, and the unbroken gauge symmetry was the 
centralizer $E_7$. In the present intersecting case, the connection embedded 
into $E_8$ is in $SU(3)$, and therefore the unbroken gauge symmetry is 
$E_6$.  The adjoint representation of $E_8$ is decomposed into 
\begin{eqnarray}
{\bf 248} &=& ({\bf 78},{\bf 1}) \oplus ({\bf 27},{\bf 3}) 
\oplus (\overline{\bf 27},\overline{\bf 3}) \oplus ({\bf 1},{\bf 8}) 
\label{decomposition}
\end{eqnarray}
as representations of the subalgebra $E_6 \times SU(3)$. 
%
%
%
Since the $E_8\times E_8$ gauge field $A_M$ has by construction 
a vev in $SU(3)$, 
the latter three gauge rotations are the moduli 
(FIG. \ref{FigE7E6}).
\begin{figure}
{\renewcommand{\arraystretch}{.5}
\begin{tabular}{cc}
\includegraphics[height=0.24\textheight]{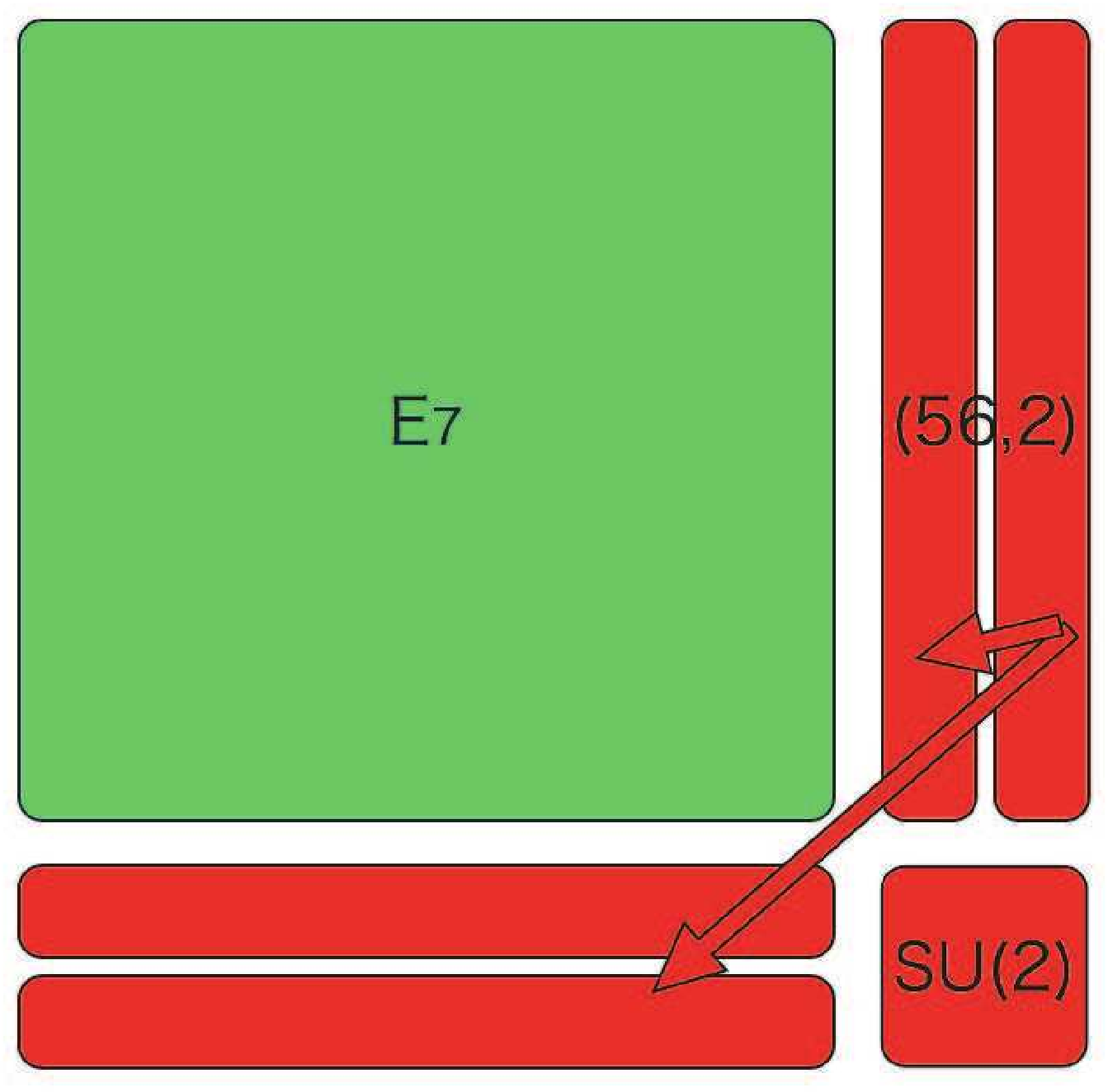}
%
& \includegraphics[height=0.24\textheight]{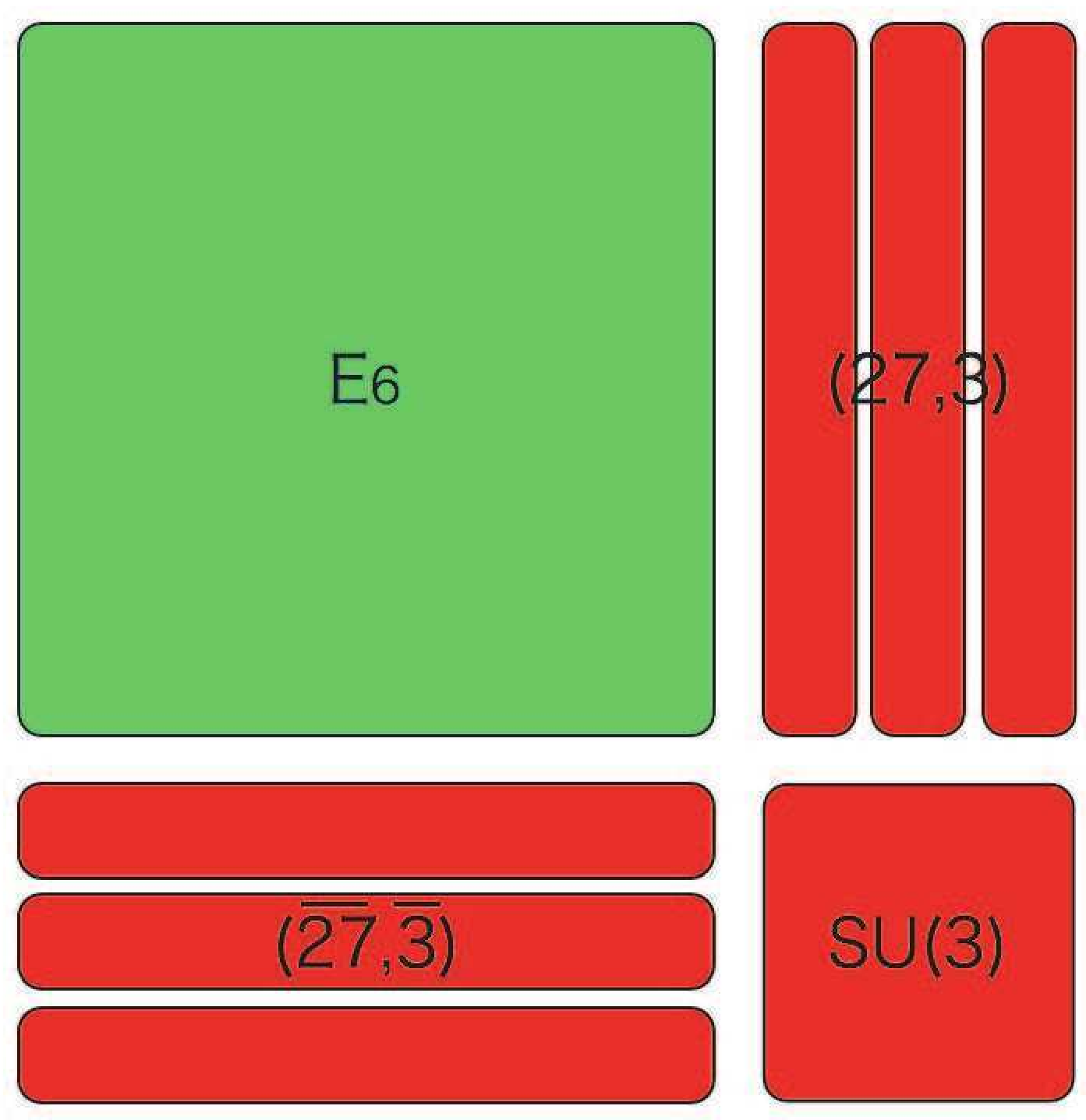}
\\
(a) & (b) 
\end{tabular}
}
\caption{\label{FigE7E6}\sl Broken generators which give rise to zero modes.
(a) The single 5-brane case. (b) The intersecting case.  
}
\end{figure}

Let us focus on the $E_6$ non-singlet moduli. As we saw in the 
symmetric 5-brane in the previous sections, spontaneously broken 
generators in $({\bf 27},{\bf 3}) \oplus
({\bf \overline{27}},{\bf \overline 3})$ give rise to Nambu-Goldstone 
bosons, each of which has one bosonic degree of freedom. 
On the other hand, since a $D=4$, ${\cal N}=1$ chiral 
supermultiplet needs {\em two} bosonic degrees of 
freedom,
the Nambu-Goldstone bosons which transform as ${\bf 27}$  
and 
${\bf \overline{27}}$ must be combined to form a single ${\cal N}=1$ 
chiral supermultiplet. 
That is, the $E_6$ non-singlet moduli form
{\em  three} chiral supermultiplets in 
the ${\bf 27}$ (or ${\bf \overline{27}}$, but not both) representation of $E_6$.


At first sight, one might think that the argument above would be contradictory  
to the well-known fact in Calabi-Yau compactifications that the number of chiral 
generations are determined by the Dirac index, in which the same decomposition 
(\ref{decomposition}) is used and  {\em one triplet} of zero modes together  
corresponds to {\em one} supermultiplet, and is not counted as three.  
Of course, it is not a contradiction, because what we consider here is not the fermionic zero 
modes of the Dirac operator, but bosonic zero modes of the gauge fields. 
As we discussed in the previous sections, they are not removed by gauge 
transformations, and necessarily exist to cancel the anomaly inflow into each of 
the two intersecting 5-branes. Each of small gauge rotation generators in 
$({\bf 27},{\bf 3}) \oplus (\overline{\bf 27},\overline{\bf 3}) \oplus ({\bf 1},{\bf 8})$ 
is an independent generator and gives rise to an independent zero mode.  
We also recall that exactly the same way of counting was done in the parallel
symmetric 5-brane case, and was indeed consistent with the index analysis 
\cite{Bellisai}.

However, it is premature to conclude that these three bosonic zero modes 
in the $({\bf 27},{\bf 3})$ representation imply three generations, because 
we have not yet examined the chiralities of their superpartners. We will do this 
in the next section. In fact, we will see that one of the three possesses the 
opposite chirality to that the other two have, and hence there is net one generation.

\subsection{\label{sec:chiral_zero_modes}Explicit computation of 
chiral zero modes
}
The ten-dimensional heterotic gaugino equations of motion reads
\begin{eqnarray}
\slash\!\!\!\!D (\omega -\frac 13 H,A)\chi -\Gamma^M \chi \partial_M \phi
+\frac 18 \Gamma^M \gamma^{AB}(F_{AB} + \hat{F}_{AB})
(\psi_M +\frac23 \Gamma_M \lambda)=0,
\end{eqnarray}
where
\begin{eqnarray}
D (\omega -\frac 13 H,A)\chi&\equiv&
\left(
\partial_M +\frac 14(\omega_M{}^{AB}-\frac13 H_M{}^{AB})\Gamma_{AB}
+ \mbox{ad}A_M
\right)\chi.
\end{eqnarray}
$\chi$ is in the adjoint {\bf 248} representation of $E_8$, and $\mbox{ad}A_M
\cdot \chi\equiv {[}A_M,\chi{]}$.
If $\psi_M=0$ and $\lambda=0$, it is simplified to
\begin{eqnarray}
\slash\!\!\!\!D (\omega -\frac 13 H,A)\chi -\Gamma^M \chi \partial_M \phi
=0.
\end{eqnarray}
Further, 
if we set $\tilde\chi \equiv e^{-\phi} \chi$, then this is equivalent
to \cite{TK0704}
\begin{eqnarray}
\slash\!\!\!\!D (\omega -\frac 13 H,A)\tilde\chi =0.
\end{eqnarray}

Since there are no nontrivial backgrounds for the four-dimensional 
$i=0,7,8,9$ directions, 
\begin{eqnarray}
\Gamma^i \partial_i \tilde{\chi} + \Gamma^\mu D_\mu(\omega -\frac 13 H,A)\tilde{\chi} =0.
\end{eqnarray}
If $\tilde{\chi}=\tilde{\chi}_{4D}\otimes \tilde{\chi}_{6D}$, the second term is 
regarded as the mass term for the four-dimensional spinor $\tilde{\chi}_{4D}$. 
We are interested in the zero modes of this Dirac operator 
$\Gamma^\mu D_\mu(\omega -\frac 13 H,A)$.

The $SO(6)$ gamma matrices in the chiral representation are 
\begin{eqnarray}
\gamma_1&=&\sigma_2 \otimes {\bf 1}\otimes {\bf 1},\nonumber\\
\gamma_2&=&\sigma_1 \otimes \sigma_1 \otimes {\bf 1},\nonumber\\
\gamma_3&=&\sigma_1 \otimes \sigma_2 \otimes {\bf 1},\nonumber\\
\gamma_4&=&\sigma_1 \otimes \sigma_3 \otimes \sigma_1,\nonumber\\
\gamma_5&=&\sigma_1 \otimes \sigma_3 \otimes \sigma_2,\nonumber\\
\gamma_6&=&\sigma_1 \otimes \sigma_3 \otimes \sigma_3.
\end{eqnarray}
The six-dimensional chiral operator is
\begin{eqnarray}
\gamma_\sharp&\equiv&-i \gamma_1 \gamma_2 
\cdots
\gamma_6 \nonumber\\
&=&\sigma_3 \otimes {\bf 1}\otimes {\bf 1}.
\end{eqnarray}
For $SO(9,1)$ gamma matrices, we take
\begin{eqnarray}
\Gamma^a&=&\gamma_{4D}^a \otimes {\bf 1}_8~~~
(a=0,7,8,9), \nonumber\\
\Gamma^\alpha&=&\gamma_{4D}^\sharp \otimes \gamma_\alpha ~~~
(\alpha=1,\ldots,6),
\end{eqnarray}
where $\gamma_{4D}^a$'s $(a=0,7,8,9)$ are the ordinary $SO(3,1)$ gamma matrices 
in the chiral representation:
\begin{eqnarray}
\gamma_{4D}^0&=&i \sigma_2 \otimes {\bf 1},\nonumber\\
\gamma_{4D}^7&=&\sigma_1 \otimes \sigma_1,\nonumber\\
\gamma_{4D}^8&=&\sigma_2 \otimes \sigma_2,\nonumber\\
\gamma_{4D}^9&=&\sigma_3 \otimes \sigma_3,\nonumber\\
\gamma_{4D}^\sharp&\equiv&-i \gamma_{4D}^0 \gamma_{4D}^7
\gamma_{4D}^8 \gamma_{4D}^9\nonumber\\
&=&\sigma_3 \otimes {\bf 1}.
\end{eqnarray}

The ten-dimensional chirality is 
\begin{eqnarray}
\Gamma_{11}&\equiv&-\Gamma^0\Gamma^7\Gamma^8\Gamma^9
\cdot \Gamma^1\cdots\Gamma^6
\nonumber\\
&=&\gamma_{4D}^\sharp \otimes \gamma_\sharp
\nonumber\\
&=&(\sigma_3\otimes {\bf 1}) \otimes (\sigma_3\otimes {\bf 1}\otimes {\bf 1}).
\end{eqnarray}

Now we consider the Dirac equation
\begin{eqnarray}
\Gamma^\mu D_\mu (\omega-\frac13 H, A)\tilde{\chi} =0.
\end{eqnarray}
%
The 16-component $SO(9,1)$ (Majorana-)Weyl spinor $\chi$ (or $\tilde{\chi}$)
is decomposed in terms of $SO(3,1)$ and $SO(6)$ spinors as
\begin{eqnarray}
{\bf 16} &=&({\bf 2}_+,{\bf 4}_+) \oplus ({\bf 2}_-,{\bf 4}_-),
\end{eqnarray}
where the subscripts are the $SO(3,1)$ and $SO(6)$ chiralities, 
$\gamma_{4D}^\sharp$ and $\gamma_\sharp$, respectively. 
Since $\tilde{\chi}$ is Majorana (but complex in this representation), 
the $({\bf 2}_+,{\bf 4}_+)$ 
and $({\bf 2}_-,{\bf 4}_-)$ components are not independent but
are transformed each other by a charge conjugation. 
%
%

As $\Gamma^\mu D_\mu (\omega-\frac13 H, A)$ is $SO(3,1)$
diagonal, it is enough to consider
\begin{eqnarray}
\gamma^\mu D_\mu (\omega-\frac13 H, A)\tilde{\chi}_{6D}&=&0,
\label{reduced_Dirac_eq}
\end{eqnarray}
with the understanding that each 
component of $\tilde{\chi}_{6D}$ is accompanied by a two-component $SO(3,1)$ Weyl spinor
with a correlated chirality ($\gamma_\sharp \gamma_{4D}^\sharp = +1$).

On the other hand, we are interested in the gaugino zero modes in $({\bf 27}, {\bf 3})$
or $(\overline{\bf 27}, \overline{\bf 3})$ in the decomposition $E_8  \supset E_6 \times SU(3)$
of ${\bf 248}$. The gauge connections $A_M$ take only nonzero values in the 
$SU(3)$ subalgebra, and we look for the zero modes $\tilde{\chi}_{6D}$ transforming as a triplet,
either ${\bf 3}$ or $\overline{\bf 3}$, of $SU(3)$.

Since $\gamma^\alpha$'s are in the form:
\begin{eqnarray}
\gamma^1&=&
\left(
\begin{array}{cc}
&-i {\bf 1}\otimes {\bf 1}\\
-i {\bf 1}\otimes {\bf 1}&
\end{array}
\right),
\nonumber\\
\gamma^{\tilde{\alpha}}&=&
\left(
\begin{array}{cc}
&\tilde{\gamma}^{\tilde{\alpha}}\\
\tilde{\gamma}^{\tilde{\alpha}}&
\end{array}
\right)~~~(\tilde{\alpha}=2,\ldots,6),
\end{eqnarray}
and $\omega_{\mu}^{~\alpha\beta}$, $H_{\mu}^{~\alpha\beta}$
and $A_{\mu}^{~\alpha\beta}$ all vanish if $\mu=1$, 
(\ref{reduced_Dirac_eq}) is reduced to two independent differential 
equations
\begin{eqnarray}
\frac ih\frac d{d x^1} {\tilde{\chi}}^{+}_{6D} + M^+ {\tilde{\chi}}^{+}_{6D} =0, \label{M+equation}\\
\frac ih\frac d{d x^1} {\tilde{\chi}}^{-}_{6D} - M^- {\tilde{\chi}}^{-}_{6D} =0,\label{M-equation}
\end{eqnarray}
where ${\tilde{\chi}}^{\pm}_{6D}$ is the upper and lower components having definite chiralities:
\begin{eqnarray}
\tilde{\chi}_{6D}=\left(
\begin{array}{c}
{\tilde{\chi}}^{+}_{6D}\\
{\tilde{\chi}}^{-}_{6D}
\end{array}
\right).
\end{eqnarray}
${\tilde{\chi}}^{+}_{6D}$ (${\tilde{\chi}}^{-}_{6D}$) is a {\bf 4} $SO(6)$ Weyl spinor, and each of the 
four components is a triplet of $SU(3)$.
Thus $M^+$ ($M^-$) is a $(4\times 3=)$ 12-by-12 matrix, given explicitly by

\begin{eqnarray}
\left(
\begin{array}{cc}
&M^-\\ M^+&
\end{array}
\right)&\equiv&\frac{h'}{h^2}
\left(
{\scriptsize
\left(
\begin{array}{llllllll}
0 & 0 & 0 & 0 & -\frac{3 i}{2} & -\frac{i}{4} & 0 & -\frac{i}{4} \\
0 & 0 & 0 & 0 & -\frac{i}{4} & -\frac{3 i}{2} & -\frac{i}{4} & 0 \\
0 & 0 & 0 & 0 & 0 & -\frac{i}{4} & -\frac{3 i}{2} & -\frac{i}{4} \\
0 & 0 & 0 & 0 & -\frac{i}{4} & 0 & -\frac{i}{4} & -\frac{3 i}{2} \\
\frac{3 i}{2} & -\frac{i}{4} & 0 & -\frac{i}{4} & 0 & 0 & 0 & 0 \\
-\frac{i}{4} & \frac{3 i}{2} & -\frac{i}{4} & 0 & 0 & 0 & 0 & 0 \\
0 & -\frac{i}{4} & \frac{3 i}{2} & -\frac{i}{4} & 0 & 0 & 0 & 0 \\
-\frac{i}{4} & 0 & -\frac{i}{4} & \frac{3 i}{2} & 0 & 0 & 0 & 0
\end{array}
\right)} 
\otimes {\bf 1}_3
\right.
\nonumber\\
&&\left.
+
{\scriptsize
\left(
\begin{array}{llllllll}
0 & 0 & 0 & 0 & -\frac{s \lambda _4}{2} & \frac{-s \lambda _1-\lambda _5}2
  & \frac{-2 \lambda _2-s \lambda_9}4 &
  0 \\
0 & 0 & 0 & 0 & \frac{\lambda _5-s \lambda _1}2 & \frac{s \lambda _4}{2} &
  0 & \frac{-2 \lambda _2-s \lambda_9}4
  \\
0 & 0 & 0 & 0 & \frac{2 \lambda _2-s \lambda_9}4 & 0 & \frac{s \lambda _4}{2} & 
\frac{s \lambda _1+\lambda_5}2 \\
0 & 0 & 0 & 0 & 0 & \frac{2 \lambda _2-s \lambda_9}4 & 
\frac{s \lambda _1-\lambda _5}2 &
 -\frac{s \lambda_4}{2} \\
-\frac{s \lambda _4}{2} & \frac{-s \lambda _1-\lambda _5}2 & \frac{
-2 \lambda _2-s \lambda_9}4& 0 & 0 & 0 & 0
  & 0 \\
\frac{\lambda _5-s \lambda _1}2& \frac{s \lambda _4}{2} & 0 & \frac{
-2 \lambda _2-s \lambda_9}4& 0 & 0 & 0 & 0
  \\
\frac{2 \lambda _2-s \lambda_9}4 & 0 &
  \frac{s \lambda _4}{2} & \frac{s \lambda _1+\lambda _5}2 & 0 & 0 & 0 & 0
  \\
0 & \frac{2 \lambda _2-s \lambda_9 }4&
  \frac{s \lambda _1-\lambda _5}2 & -\frac{s \lambda _4}{2} & 0 & 0 & 0 & 0
\end{array}
\right)
}
\right),\nonumber\\
\end{eqnarray}
where 
$\lambda_9\equiv 3 \lambda _3+\sqrt{3}\lambda_8$. 
In identifying the spin connection as an $SU(3)$ gauge connection, 
$s=\left(
\renewcommand{\arraystretch}{.7}
\begin{array}{cc}
&~1\\-1&
\end{array}
\right)$ can either be mapped to $i$, or to $-i$, 
and depending on this choice, the $SU(3)$ gauge connection 
matrix becomes one
in the {\bf 3} representation, or in the $\overline{\bf 3}$ representation.

As we already mentioned, ${\tilde{\chi}}^{+}_{6D} $ and ${\tilde{\chi}}^{-}_{6D} $ 
are not independent; we have only to solve the equation (\ref{M+equation}),
and the solutions to (\ref{M-equation}) may then be obtained by a charge 
conjugation. To solve (\ref{M+equation}), we diagonalize $M^+$ to obtain 
its eigenvalues. Let $i\lambda$ be an eigenvalue of the {\em constant} matrix 
$\left(\frac {h'}{h^2}\right)^{-1}M^+$, and $\psi_\lambda(x^1)$ 
be the corresponding eigenfunction, then they satisfy
\begin{eqnarray}
\frac i{h} \psi'_\lambda + i\lambda \frac{h'}{h^2}\psi_\lambda&=&0.
\end{eqnarray}
This is solved to give
\begin{eqnarray}
\psi_\lambda(x^1)&=&\mbox{const.} (h(x^1))^{-\lambda}.
\end{eqnarray}
Thus, for each eigenvalue,  there exists a zero mode of the Dirac operator.
Since $\xi$ is negative for positive tension, if $\lambda<1$, the mode is 
localized near $x^1=0$, while if $\lambda \geq 1$, it is not localized, being 
either non-normalizable or localized rather at ``infinity'' $x^1=\pm\frac{h_0}{|\xi|}$.

The list of eigenvalues of $\left(\frac {h'}{h^2}\right)^{-1}M^+$ is as follows:
If $s=+i$, the eigenvalues are
\begin{eqnarray}
\left\{2 i,\frac{3 i}{2},\frac{3 i}{2},i,-i,i,i,i,\frac{3 i}{2},\frac{3 i}{2},\frac{7
  i}{2},\frac{7 i}{2}\right\},
\label{eigenvalues+i}
\end{eqnarray}
while
if $s=-i$, they are
\begin{eqnarray}
\left\{2 i,\frac{3 i}{2},\frac{3 i}{2},i,2 i,4 i,2 i,2 i,-\frac{i}{2},-\frac{i}{2},\frac{3
  i}{2},\frac{3 i}{2}\right\}.
\label{eigenvalues-i}
\end{eqnarray}

We can clearly see an asymmetry between (\ref{eigenvalues+i}) and 
(\ref{eigenvalues-i}), in particular that the former has only one negative 
(times imaginary unit) eigenvalue, while the latter has two negative 
eigenvalues. 
Assuming that the branes have positive tension so that the function $h(x)$ 
has the profile shown in FIG. \ref{negative_xi}, 
these are the only modes whose profiles 
have a peak at $x^1=0$ or $z=0$ in the coordinate (\ref{z}). 
The same is also true for the original gaugino variable $\chi = h \tilde{\chi}$
(although the modes with $\lambda=+1$ then become constant).
This result implies that there are indeed three localized modes, and two of 
them are in one (say,  ({\bf 27},{\bf 3})) representation, and the rest belongs to 
the other (($\overline{\bf 27},\overline{\bf 3}$)) representation.


\section{Conclusions and Discussion}
In this paper, we have shown that 
there exist three localized zero modes as $D=4$, ${\cal N} =1$ 
supermultiplets on the system of two intersecting 5-branes in the 
$E_8 \times E_8$ heterotic string theory. 
By using the standard embedding in the known smeared solution, 
we have constructed a heterotic background 
and 
explicitly solved the Dirac equation on this background. We have found that 
two of them are in the {\bf 27} representation of $E_6$, 
and one in the $\overline{\bf 27}$ representation. They give rise to net {\em one} 
{\bf 27} of massless chiral fermions in the four-dimensional spacetime. 
This is the first example of a brane set-up in heterotic string theory  
that supports, after compactifying some of the transverse dimensions,
four-dimensional chiral matter fermions transforming as an $E_6$ gauge 
multiplet.

Intuitively, the chirality flip of one of the three zero modes can be understood 
as follows: the further one goes away from the intersection to the $x^3$ or $x^4$ 
direction along one 5-brane,  the less one feels the effect of the other brane, and in the end one would observe as if there were only a single 
symmetric 5-brane. The gauge connection then becomes smaller than $SU(3)$, 
and approaches to $SU(2)$. As we have seen in the previous sections, the zero modes 
on a single 5-brane are 30 six-dimensional supermultiplets, which are of course 
nonchiral as four-dimensional supermultiplets upon a dimensional reduction. 
They are regarded as two of the three columns and rows shown FIG. \ref{FigE7E6}(b),
and have opposite chiralities. Similarly, if one goes away from the intersection 
to the $x^5$ or $x^6$ direction, one will observe a reduction of the gauge connection 
from $SU(3)$ to a different $SU(2)$, and will see, again,  a pair of nonchiral zero modes 
which correspond to different pair of columns and rows in FIG. \ref{FigE7E6}(b).
Therefore, since there are only three sets of zero modes, the chirality 
of one of them must be opposite to that of the other two.

It is worth mentioning that this chirality flip also agrees with the
analysis of K\"{a}hler coset sigma models \cite{IKK}. In general, the dynamics of 
Nambu-Goldstone modes is described by a low-energy sigma model action 
constructed as a group coset associated with the corresponding 
spontaneously broken symmetries. In the ${\cal N}=1$ supersymmetric case, 
the target space must be K\"{a}hler. $E_8/[E_6\times SU(3)]$ is not a K\"{a}hler 
coset; no wonder because this is not the moduli space of the 
intersecting 5-branes (since the adjoint of $SU(3)$ also belongs to the moduli). 
On the other hand, there {\em are}  K\"{a}hler cosets 
which contain three ${\bf 27}$'s of $E_6$. They are $E_8/[E_6\times SU(2)\times U(1)]$ 
and $E_8/[E_6\times U(1)^2]$.
It turns out that, in both cases, the chirality of one of three supermultiplets are 
opposite to the other two\footnote{We thank T.~Kugo, H.~Kunitomo and N.~Ohta 
for discussions on this point.}.
Although neither of them coincides exactly with the moduli space of the 
intersecting 5-branes, this is just what we have encountered in the present 
analysis and may be regarded at least as a suggestive fact.

It will be extremely interesting if this set-up could be used to realize 
the $E_6$ grand unification scenario \cite{E6} by using branes \cite{Kawamura} 
in string theory. For this purpose, 
we need to generalize it to a more realistic brane system which supports 
three generations. In principle, one could do this by replacing one of the single
5-brane with three 5-branes and consider the intersection with the other 5-brane.
This is also suggested by the study of duality between the orbifolded or 
generalized conifold and a system of intersecting NS5-branes \cite{Aganagic_et_al}. 
We hope to report on this issue in the near future.

\begin{acknowledgments}
We would like to thank 
Tohru Eguchi, Satoshi Iso, Hikaru Kawai, Taichiro Kugo, Hiroshi Kunitomo, 
Nobuyoshi Ohta and Shigeki Sugimoto for illuminating discussions.
We are also grateful to 
Keiichi Akama, 
Masafumi Fukuma,
Machiko Hatsuda,
Takeo Inami, 
Akihiro Ishibashi, 
Katsushi Ito, 
Katsumi Itoh,
Hiroshi Itoyama,
Yoichi Kazama,
Yoshio Kikukawa,
Yoshihisa Kitazawa, 
Hideo Kodama, 
Nobuhiro Maekawa, 
Nobuhito Maru,  
Yoji Michishita, 
Muneto Nitta,
Kazutoshi Ohta, 
Soo-Jong Rey,
Tomohiko Takahashi, 
Shinya Tomizawa, 
Tamiaki Yoneya
and 
Kentaro Yoshida
for discussions and comments.
We thank YITP for hospitality and support 
during the workshop: ``Branes,
Strings and Black Hole'', where part of this work was done. 
This work is supported by 
Grant-in-Aid
for Scientific Research (C) \#20540287-H20 from
The Ministry of Education, Culture, Sports, Science
and Technology of Japan.

\end{acknowledgments}


\end{document}